\documentclass[aps,prl,amsmath,twocolumn,amssymb,titlepage]{revtex4-1}

\usepackage{graphicx}
\usepackage{nicefrac}
\usepackage{amsfonts}

\usepackage{multirow} 
\usepackage{tabularx}

\usepackage{bm}
\usepackage{hyperref}

\begin{document}
\title{First-principles investigation of MnP magnetic states precursors of superconductivity under high pressure}
\author{Pietro Bonf\`a}
\email{pietro.bonfa@fis.unipr.it}
\affiliation{Department of Physics and Earth Sciences, University of Parma, Italy}
\author{Ifeanyi John Onuorah}
\affiliation{Department of Physics and Earth Sciences, University of Parma, Italy}
\author{Roberto De Renzi}
\affiliation{Department of Physics and Earth Sciences, University of Parma, Italy}

\begin{abstract}
The discovery of a superconducting dome in the proximity of the magnetic to paramagnetic transition in the electronic phase diagram of MnP as a function of hydrostatic pressure has renewed the interest in the magnetic properties of this binary pnictide. 
Here we present an \emph{ab initio} study obtained with Density Functional Theory (DFT)
simulations as a function of pressure. We show that the itinerant-electron magnetism of MnP is well characterized by the mean-field Kohn-Sham method which correctly describes the ambient pressure magnetic states, the anomalous trend of the lattice parameters as a function of pressure and the critical pressure for the disappearance of the magnetic order. We are finally able to confirm the nature of the new helical structure observed at high pressure.
\end{abstract}

\maketitle

Manganese phosphide (MnP) has continuously attracted both theoretical and experimental attention for decades. 
More precisely, its magnetic properties have been studied to clarify the relevance of Lifshitz critical points for this material, to address its magnetocaloric effect and, recently, the role of magnetic excitations in promoting the appearance of superconductivity (SC) under hydrostatic pressure \cite{Moon_Cable_Shapira_1981,PhysRevLett.44.1692,PhysRevLett.114.117001,1.3072022}.
Indeed, the presence of a SC dome with a maximum $T_{\mathrm{C}}\sim $~1~K at the edge of the magnetic-to-paramagnetic transition resembles the electronic phase diagram of many unconventional high temperature superconductors. 
Whether this newly discovered Mn based superconductivity is unconventional or not is
still debated. For instance recent \emph{ab initio} simulations \cite{Chong2016} predict, within the
framework of weak coupling BCS, a superconducting transition temperature in
agreement with the experiment.

\begin{figure}
\includegraphics[]{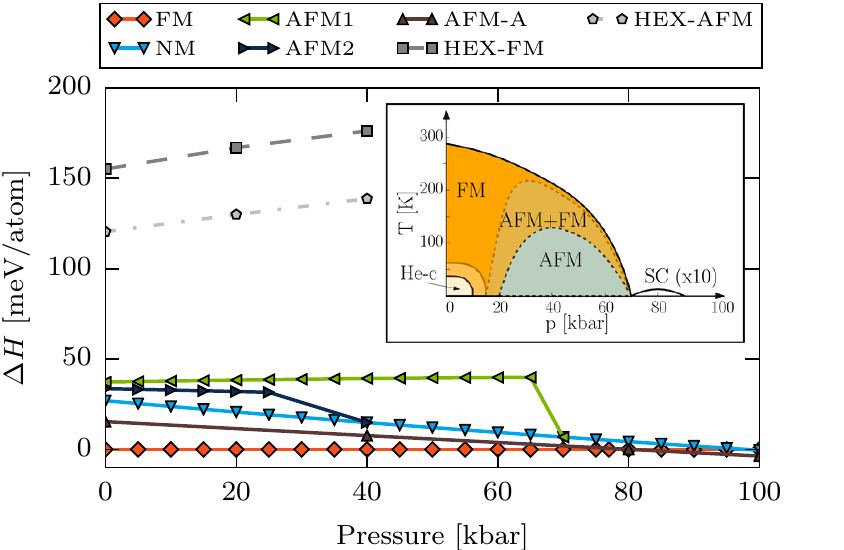}    
\caption{(Color online) Enthalpy difference with respect to the FM state as a function of applied pressure. 
NM refers to the non magnetic (spinless) simulations while the various AFM orders are discussed in Tab.~\ref{tab:afmorders} and in the text. The inset sketches the electronic phase diagram as obtained from Refs.~\onlinecite{1511.09152,1603.03367,PhysRevB.93.100405,PhysRevLett.114.117001}. The shaded areas tentatively identify the regions where two different magnetic states coexist.
}\label{fig:allEnthalpies}
\end{figure}

The magnetic part of the phase diagram, sketched in the inset of Fig.~\ref{fig:allEnthalpies}, has been extensively discussed in the literature \cite{PhysRev.135.A1033,1.1708333,PhysRevLett.114.117001,PhysRevB.93.100405}. 
MnP crystallizes in $Pnma$ (62) symmetry with a 4 Mn atoms unit-cell having lattice constants $a=5.2361$~\AA, $b=3.1807$~{\AA} and $c=5.8959${\AA} which stems from a distortion of the 2 Mn atoms NiAs-type hexagonal (HEX) cell \cite{1511.09152,Hulliger1968}.
Below the paramagnetic (PM) to ferromagnetic (FM) transition at $T_{\mathrm{C}}=291$~K, Mn $3d$ orbitals have a reduced moment of about 1.3 $\mu_{\mathrm{B}}$ lying parallel to the $b$ axis \cite{1.1708333}. 
Below 50~K, a complex double-helix order with propagation vector $\mathbf{q}=(0,0,0.117)$ develops \cite{1.1708333,JPSJ.83.054711}. 
A hydrostatic pressure of about 20~kbar induces the transition from the double-helix structure to a different type of antiferromagnetic 
(zero macroscopic magnetization) ground state with $T_{\mathrm{N}} \sim 150$~K (labeled AFM in the inset of Fig.~\ref{fig:allEnthalpies}).
All these transitions are characterized by the presence of regions of coexistence of FM and AFM phases observed both as a function of temperature and pressure \cite{1603.03367}.

The nature of the new magnetic order which precedes the appearance of superconductivity is still puzzling.
Two possible scenarios have been obtained from x-ray, neutron scattering and $\mu$SR experiments.
The first \cite{1511.09152} identifies the new state still as a double helix, albeit with an
 increased propagation vector from the value $|\mathbf{q}|=0.11$ observed at ambient pressure to the value $|\mathbf{q}|\sim 0.25$. 
The second \cite{1603.03367,PhysRevB.93.100405} discusses a transition of the propagation vector from $\mathbf{q} \parallel \hat{c}$ to $\mathbf{q} \parallel \hat{b}$.

To clarify the magnetic properties of this material we have accurately refined first principle DFT simulations.
To date, \emph{ab initio} methods have been used to asses the presence of antiferromagnetic (AFM) collinear spin structures and phonon mediated superconductivity \cite{Chong2016,PhysRevB.81.224426}.
In this letter we first discuss the possible magnetic orders of MnP considering the conventional collinear spin density and a full relaxation of the unit cell and then refine the predictions by using the non-collinear spin spiral formalism to describe the helical magnetic orders as a function of applied pressure.
One may argue that the KS mean field approach is generally not suitable for describing the strongly correlated $d$ orbitals. However, the generalized gradient approximation (GGA) is found to be surprisingly accurate in accounting for the evolution of the magnetic properties of the MnP probably owing to the itinerant nature of its magnetic states \cite{PhysRevB.64.085204,PhysRevB.69.132411}.

The electronic properties of MnP have been analyzed with both pseudopotential and full-potential approaches \cite{QE-2009,elk,PhysRevLett.78.1396,PhysRevB.40.3616,PhysRevB.13.5188,sssp,Lejaeghereaad3000,PhysRevB.50.17953}. The reader is referred to the supplementary material for the computational details.
The magnetic states of MnP under pressure were first examined in the collinear spin formalism. 
As already discussed in Ref.~\onlinecite{PhysRevB.81.224426}, there are three collinear antiferromagnetic orders that can be realized in the orthorhombic unit cell.
These are reported for the reader's convenience in Tab.~\ref{tab:afmorders} with the same labels used in Ref.~\onlinecite{PhysRevB.81.224426}. 
We additionally considered an A-type antiferromagnetic order along the $b$ axis (labeled AFM-A in Tab~\ref{tab:afmorders}). 

The total enthalpy as a function of the applied pressure for the various magnetic configurations of both the hexagonal and the orthorhombic lattice structures is shown in Fig. \ref{fig:allEnthalpies}.

The stable collinear phase at ambient pressure is the orthorhombic crystal structure with a FM ground state and 1.4 $\mu_{\mathrm{B}}$ per formula unit (f.u.), a value that is in excellent agreement with the experiment \footnote{The FM state is the closest collinear equivalent of the helical state which is observed below 50K at ambient pressure.}.
The AFM NiAs-type structure, from which the distorted MnP cell originates, has a much larger energy and it is further suppressed with applied hydrostatic pressure. This compares well with the evidence of an increasing departure from the hexagonal structure obtained from the experimental evolution of the lattice parameters with pressure \cite{1511.09152}.

In agreement with previously published results \cite{PhysRevB.81.224426}, none of the AFM1-3 orders is stabilized by increasing the pressure. We find that, by relaxing the lattice structure, the AFM3 order is suppressed even at ambient pressure. This is expected since the results of Ref.~\onlinecite{PhysRevB.81.224426} show that the staggered moment on Mn for AFM3 quickly drops to zero immediately after the volume of the unit cell is reduced.
For the other two magnetic orders, AFM1 and AFM2, the enthalpy difference with respect to the ferromagnetic state slightly increases or remains constant as a function of pressure and therefore they cannot represent the ground state at high pressure.

\begin{table}
    \begin{tabular}{|c|c|c|c|c|c|}
    \hline 
    Mn atoms & Mn$_{1}$ & Mn$_{2}$ & Mn$_{3}$ & Mn$_{4}$ & Mn$_{4'}^{b}$ \tabularnewline
    \hline
    Interatomic & \multicolumn{2}{c|}{$d_{1}$} & \multicolumn{2}{c|}{$d_{1}$} & \tabularnewline
    \cline{2-6} 
    distance &  & \multicolumn{2}{c|}{$d_{2}$} & \multicolumn{2}{c|}{$d_{3}$} \tabularnewline
    \hline 
    AFM1 & $\Longrightarrow$ & $\Longrightarrow$ & $\Longleftarrow$ & $\Longleftarrow$ & $\Longleftarrow$\tabularnewline
    \hline 
    AFM2 & $\Longrightarrow$ & $\Longleftarrow$ & $\Longleftarrow$ & $\Longrightarrow$ & $\Longrightarrow$\tabularnewline
    \hline 
    AFM3 & $\Longrightarrow$ & $\Longleftarrow$ & $\Longrightarrow$ & $\Longleftarrow$ & $\Longleftarrow$\tabularnewline
    \hline 
    AFM-A & $\Longrightarrow$ & $\Longrightarrow$ & $\Longrightarrow$ & $\Longrightarrow$ & $\Longleftarrow$ \tabularnewline
    \hline
    \end{tabular}

    \caption{Possible collinear antiferromagnetic configurations of Mn$_{1-4}$ atoms in a single unit cell and the nearest neighbor Mn$_{4'}^{b}$ along $+\mathbf{b}$, with their relation to the inter-Mn distances $d_{1}<d_{2}<d_{3}$. The first three rows identify the collinear antiferromagnetic orders  discussed in Ref.~\onlinecite{PhysRevB.81.224426}. The last row represents the A-type antiferromagnet along $\mathbf{b}$.}\label{tab:afmorders}
\end{table}

\begin{figure}
\includegraphics[]{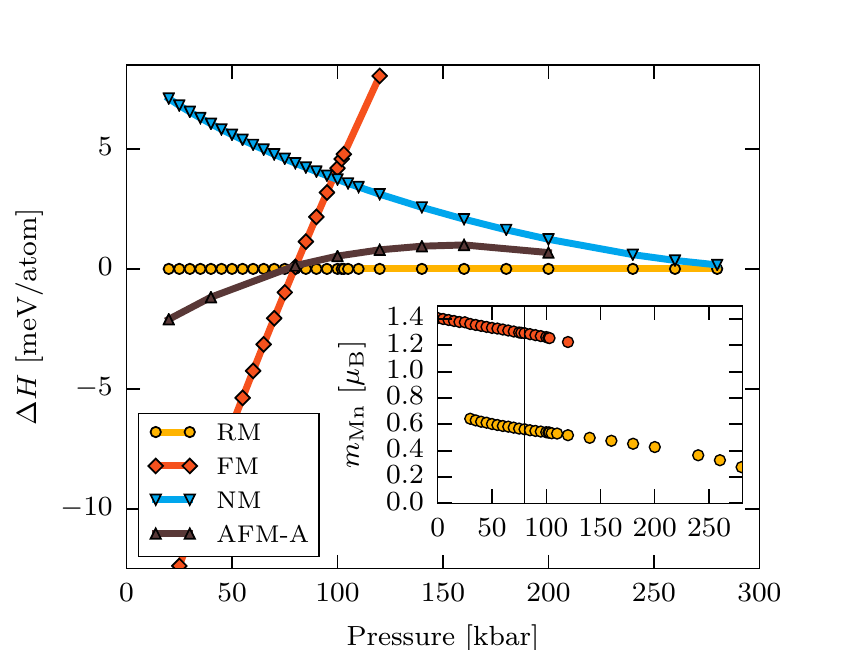}
\caption{(Color online) Enthalpy difference with respect to the RM state. The inset shows the comparison between the full 1.4 $\mu_{\mathrm{B}}$/f.u. of the ferromagnetic ground state and the reduced moment of about 0.6 $\mu_{\mathrm{B}}$/f.u. of the RM ground state. }\label{fig:selectedEnthalpies}
\end{figure}

\begin{figure*}
\includegraphics[]{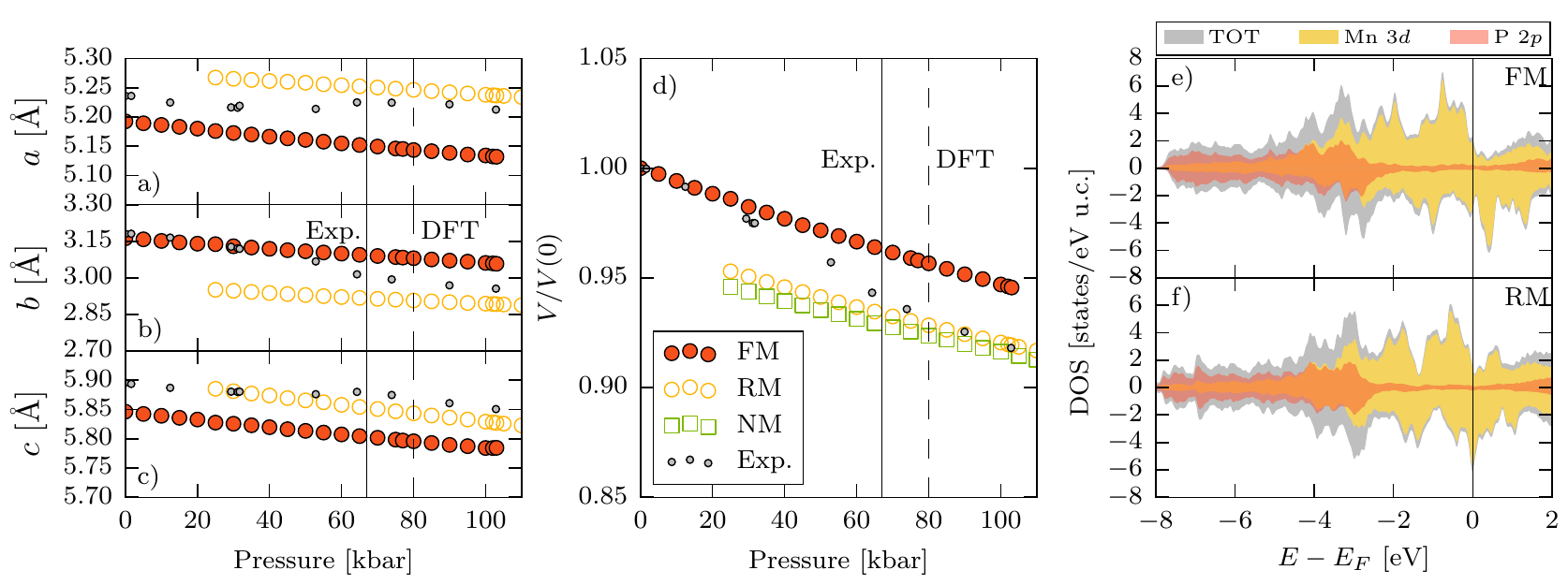}
\caption{(Color online) Panels a)-d): evolution of the lattice parameters and of the volume of the unit cell as obtained from \emph{ab initio} results compared with the experimental results from Ref.~\onlinecite{1511.09152}. In panel d), the small inaccuracy in the estimation of the unit cell volume by DFT makes the direct comparison of the two structures with the experimental results not very effective, but the comparison of the relative contraction of the experimental volume and of the first principles estimations perfectly match.
Panels e) and f): density of states for the FM and the RM  ground states at $p=120$~kbar.}\label{fig:lattice}
\end{figure*}


Zooming into Fig.~\ref{fig:allEnthalpies} one may notice that the FM state ceases to be the ground state above $p=80$~kbar.
As shown in Fig.~\ref{fig:selectedEnthalpies}, in this range our simulations find an unexpected new ferromagnetic phase with reduced Mn moment $m_{\mathrm{Mn}} \sim 0.6 \mu_{\mathrm{B}}$, that we label RM. The result is shown in Fig.~\ref{fig:selectedEnthalpies} as the enthalpy difference between several of the considered states and the new RM  state, that remains energetically favored up to 280 kbar. Only then, the spinless (NM) electron density becomes the ground state \footnote{We cannot exclude the possibility of a phase transition to a lattice structure with lower symmetry at $p<280$~kbar}.
Interestingly, Ref.~\onlinecite{PhysRevB.93.100405} identifies a ferromagnetic phase at high temperature with a reduced moment of 0.7 $\mu_{\mathrm{B}}$ already at $p$=20~kbar. Our results are strictly valid at zero temperature so the relation between DFT and experiment is only indirect.

The transition to the RM state has a strong effect on the lattice parameters with the most prominent outcome being a substantial reduction of $b$ accompanied by a small increase of both $a$ and $c$ again in good agreement with the experimental observations.
For completeness we mention that the AFM-A magnetic state displays a similar contraction of the $b$ axis and reduced Mn magnetic moment. In addition, it has  much smaller enthalpy than AFM1-3 orders. 
However, the AFM-A state is never found to be the ground state, even if, at p=80~kbar, it is almost degenerate with the FM and the RM states.

The evolution of the lattice parameters with applied pressure is presented in Fig.~\ref{fig:lattice}a-c. The general agreement of the GGA estimation of the cell size is within 2\% of the experiment. While the $b$ axis is very well reproduced, the $a$ axis is underestimated and the $c$ axis is overestimated. This gives a unit cell volume 96.00~\AA$^{3}$ to be compared with the experimental result of 98.89~\AA$^{3}$ \cite{1511.09152}. The trend of the lattice constants is non-monotonic. It is characterized by the elongation of the $a$ and the $c$ axis at the expenses of $b$ at about 50~kbar \cite{1511.09152}.
This trend follows the expected evolution for a transition from the FM to the RM state as shown by Fig.~\ref{fig:lattice}a-c.
However, a similar evolution is predicted also for the FM to NM transition as evidenced by the volume trend displayed in Fig.~\ref{fig:lattice}d (see also Fig.~\ref{fig:extendedlatpar} in supplemental material).

A comparison of the total and the projected density of states (DOS) for the RM and the FM states at $p=120$~kbar is presented in Fig.~\ref{fig:lattice}e and \ref{fig:lattice}f.
In the FM state, a spin polarization of $1.33 \mu_{\mathrm{B}}$ and $-0.11 \mu_{\mathrm{B}}$ for Mn $3d$ and P $2p$ is respectively obtained.
In the RM state, the polarization of the $3d$ and $2p$ orbitals are reduced to $m_{Mn} = 0.62 \mu_{\mathrm{B}}$ and $m_{P} = -0.05 \mu_{\mathrm{B}}$.
The dominant contribution to the DOS at the Fermi level $E_{F}$ stems from Mn $3d$ electrons in both the FM and the RM states. The hybridization with the $2p$ P orbitals gives rise to a series of features which are localized around -3~eV in agreement with previous photoemission spectroscopy results \cite{PhysRevB.69.132411}.
The main features of the $3d$ states in the spin-up channel are similar in the FM and the RM states. 
On the other hand, a marked difference in the $3d$ spin-down channel is observed. The two peaks at -1.5 eV in the FM state are shifted toward slightly lower energies and a substantial increasing number of minority spin states at the Fermi energy is observed owing to the collapse of the sharp features around -0.5 and 1.5 eV into a broad structure around 0 eV in the RM state. The resulting increase of the density of states at the Fermi level could play a crucial role 
in stabilizing a finite superconducting critical temperature in the BCS framework.

Let us summarize the results of our collinear calculations. Although the formalism may not be sufficient to fully describe the magnetic order of the system, it correctly predicts the evolution of the lattice parameters, the details of the FM phase, including the features of the density of states, and the presence of a critical pressure $p_{c} \sim 80$~kbar.
Above $p_{c}$, the identification of a new reduced-moment, compressed-volume phase, significantly arising near the FM to PM transition where superconductivity emerges, is an additional proof of the presence of competing magnetic states.

\begin{figure*}
\includegraphics[]{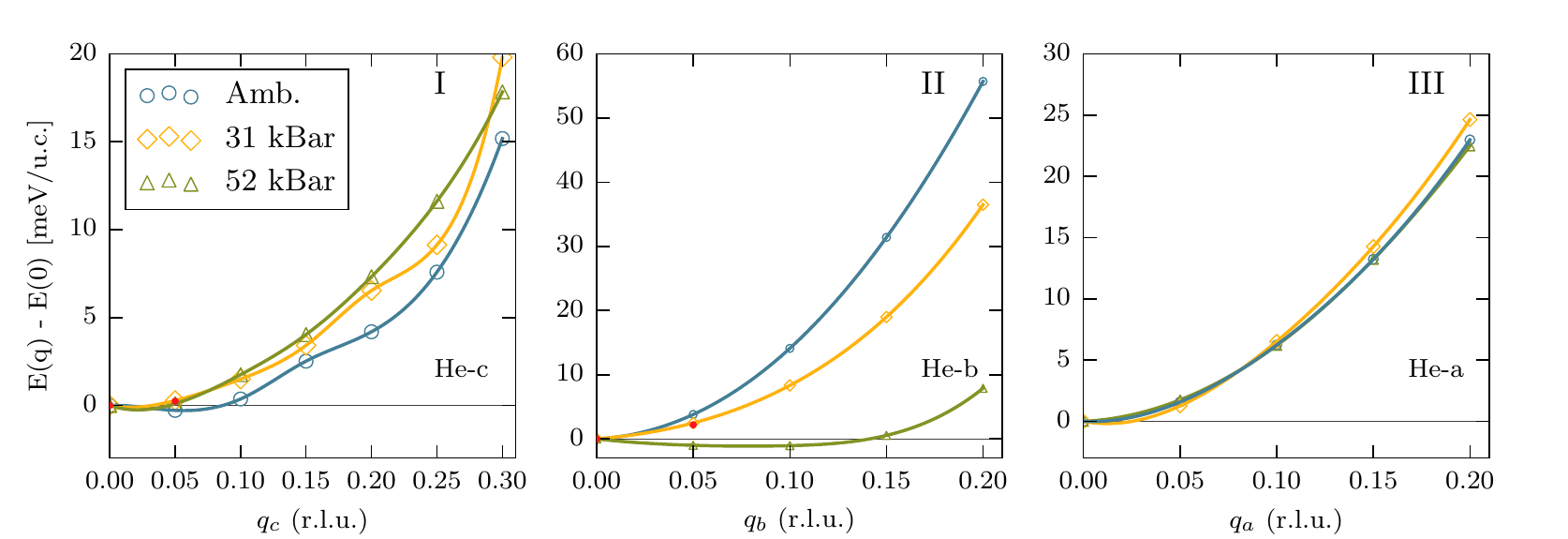}
\caption{(Color online) Energy differences for the three helical states with propagation vectors along $a$, $b$ and $c$ at 0 kbar, p=31~kbar and p=52~kbar. The size of the symbols represents an estimate of the accuracy of the simulation. The lines are guide for the eyes.
To verify the impact of the optimization of the atomic positions on the results we performed a series of simulations with the atoms fixed in ambient pressure positions. The results, presented as small red dots, always fall on top of the previous within the accuracy of the simulation.}\label{fig:spirals}
\end{figure*}

Let us now go back to the experimental magnetic phase diagram of Fig.~\ref{fig:allEnthalpies} (inset) which is characterized by a series of non collinear and non commensurate structures, extremely sensitive to applied pressure.
Contrasting results regarding the nature of the long range AFM order stabilized above $p$=20~kbar have been published \cite{1511.09152,PhysRevB.93.100405}.
The authors of Ref.~\onlinecite{1511.09152} describe an almost commensurate propagation vector $\mathbf{q} \sim (0, 0, 0.25)$ with the same form factor of the double helical state seen at ambient pressure.
On the other hand, very recent neutron scattering and $\mu$SR experiments suggest the transition to a different magnetic order with propagation vector along $\mathbf{b}$ \cite{PhysRevB.93.100405,1603.03367}.

In order to discern the true nature of the new AFM phase, we adopted the spin spiral formalism within a full-potential description.
The spin excitations of MnP have been successfully discussed as the result of competing exchange interactions between neighboring manganese \cite{Yano201333}.
As it will become clear, the shrinking of the $b$ lattice constant plays therefore a crucial role in determining the prevalence of one interaction over the other \footnote{To obtain accurate results, we decided to fix the lattice parameters to the known experimental values\cite{1511.09152}.
Since, to the best of our knowledge, the atomic positions as a function of the pressure are still unknown, we optimized them in the ferromagnetic state including also the spin orbit coupling.}.

For the sake of clarity we label the various helical states with the lattice parameters along which the helix propagates.
The general propagation vector can be written, in relative coordinates, as $\mathbf{q}=(q_{a},q_{b},q_{c})$. We explored the stability of the three helices He-a, He-b and He-c by varying separately the values of $q_{a},q_{b}$ and $q_{c}$ between 0 and 0.2 at ambient pressure, p=31~kbar and p=52~kbar. For the He-c case, in order to compare with the experimental findings of Ref.~\onlinecite{1511.09152}, we explored the stability of helical states up to $q_{c}=0.3$.
The results are reported in Fig.~\ref{fig:spirals}. 
 
At ambient pressure, the energies of He-c with $q_{c}<0.1$ are degenerate within the accuracy of our simulation (see Fig.~\ref{fig:spirals}I. This suggests the presence of an instability towards the He-c order that is indeed observed experimentally. 
We also notice the presence of a variation of the slope in the energy dependence on $q_{c}$. This  feature, which is below the accuracy of our present simulations, may deserve further investigation owing to the observation of a magnetic peak with $\mathbf{q} \sim (0,0,0.25)$ \onlinecite{1511.09152}.
However, with increasing pressure, the instability seemingly disappears and we therefore conclude that He-c is not the high-pressure helical state.

Fig.~\ref{fig:spirals}II shows the trend for He-b as a function of applied pressure. Compared with Fig.~\ref{fig:spirals}I, a steep increase of the energy difference with respect to the FM state as a function of $q_{b}$ is observed at ambient pressure (blue curves, notice the different vertical scales of Fig.~\ref{fig:spirals}). The same holds for the results obtained at $p=$31~kbar. 
However, it is immediately apparent that this spiral order is strongly dependent on the applied pressure.
Indeed, at $p=$52~kbar, the helices with $0 \leq q_{b} \leq 0.15 $ become
nearly degenerate in energy.

These results thus strongly support the conclusion that He-$b$ is the AFM order which precedes the superconducting state in agreement with Ref.~\onlinecite{PhysRevB.93.100405}.
Our findings suggest a critical pressure for stabilizing the new phase larger by a factor two compared to the experiment.
However, given the substantial impact of the unit cell volume on the magnetic structure, it is
important to underline that we utilize lattice parameters taken from experiments in
the same pressure range where $\mu$SR results \cite{1511.09152} identify a coexistence of regular FM
and helical states. Hence probably the experiment measures an average
lattice. The true lattice parameters, and especially $b$, in the helical state may be even smaller and with such smaller values DFT would predict a lower critical pressure.

For completeness, the magnetic order with propagation vector along $a$ is presented in Fig.~\ref{fig:spirals}III.
The variation of the total energy differences as a function of $q_{a}$ is monotonic and it does not present instabilities towards any finite $q_{a}>0$.
For the three considered pressures, the variations are mostly within the accuracy of the simulation and we do not expect a significant effect of the applied pressure.

In summary, we present first-principle simulations aimed at assessing the true magnetic state of MnP under hydrostatic pressure relevant for the emergence of superconductivity.
Our result are consistent with the evolution of the lattice parameters observed experimentally.
We identify a RM phase which, together with the AFM-A order, becomes degenerate with the FM state at a critical pressure $p_{c}$=80~kbar.
These facts support the presence of competing magnetic states in the vicinity of superconductivity and suggests that critical spin fluctuations may have a significant role in stabilizing the superconducting state although this point would clearly require further investigations with more refined theoretical methods.

Importantly, we are able to reproduce both the low temperature He-$c$ and the FM state experimentally observed at ambient pressure. We also confirmed that the He-$b$ replaces the He-$c$ as the ground state precursor of the superconducting dome.

Finally, it is worth stressing that our quantitatively accurate predictions support the use of the DFT mean field approach (within GGA) as a valuable strategy for further investigation of the electronic properties of this material.

We acknowledge Yishu Wang and his coauthors \cite{1511.09152} for sharing their data on the evolution of the lattice constants with applied pressure. 
We thank Lars Nordstr\"om and Gianni Profeta for fruitful discussion.
We are grateful for the computing resources provided by CINECA within the Scientific Computational Project CINECA ISCRA Class C (Award HP10C5EHG5, 2015) and STFC's Scientific Computing Department.
This work was supported by the grants PRIN project 2012X3YFZ2 and from the European Union Horizon 2020 research and innovation programme under grant agreement No 654000.

\section*{Supplementary Information}

\setcounter{equation}{0} \setcounter{figure}{0}

\renewcommand{\theequation}{S\arabic{equation}}
\renewcommand{\thefigure}{S\arabic{figure}}

\subsection{Computational details}

The electronic properties of MnP have been analyzed with both pseudopotential and full-potential approaches.
\begin{figure}[h]
    \includegraphics[]{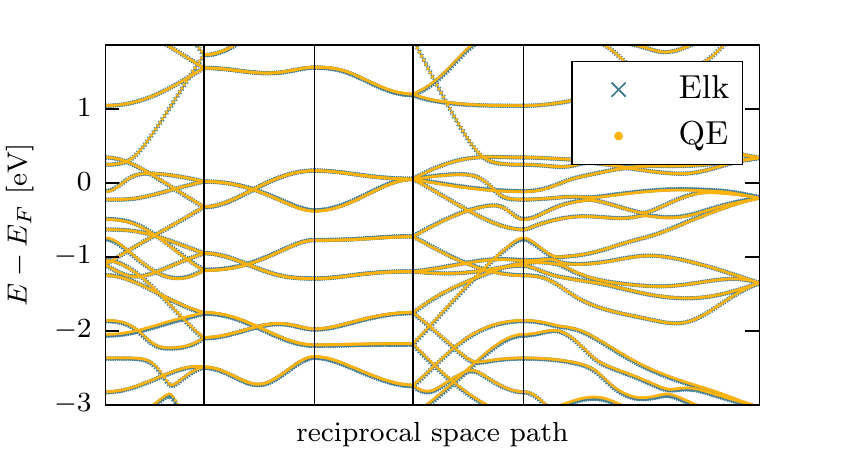}
    \caption{(Color online) Comparison of the band structures obatined with the planewave and pseudopotential approach (QE) and the LAPW basis (Elk). }\label{fig:bands}
\end{figure}

In both cases we used the generalized gradient approximation (GGA) as parametrized by Perdew, Burke and Ernzerhof \cite{PhysRevLett.78.1396}.
For the pseudopotential based simulations, the Kohn and Shan orbitals have been expanded in a plane-wave basis and the core wavefunction has been reconstructed with the projector augmented wave method using the \textsc{QuantumESPRESSO} (QE) suite of codes \cite{QE-2009,sssp,Lejaeghereaad3000,PhysRevB.50.17953}.
In order to achieve convergence on the absolute value of the estimated pressure, a large basis set must be used. We set the kinetic energy and charge cutoffs to 90~Ry and 900~Ry respectively. A Monkhorst-Pack (MP) mesh of $8 \times 8 \times 8$ is used to sample the reciprocal space for plane-wave simulations \cite{PhysRevB.13.5188}. These parameter provide total energies converged to 0.5 meV/atom.
The thresholds for the relaxation of the lattice parameters and the forces were set to 1 mRy/a.u. and to less that 0.5~kbar from target pressure.
The enthalpy for the various phases is obtained by considering the optimized volume and total energy and the target pressure. 

\begin{figure}
    \includegraphics[scale=0.5]{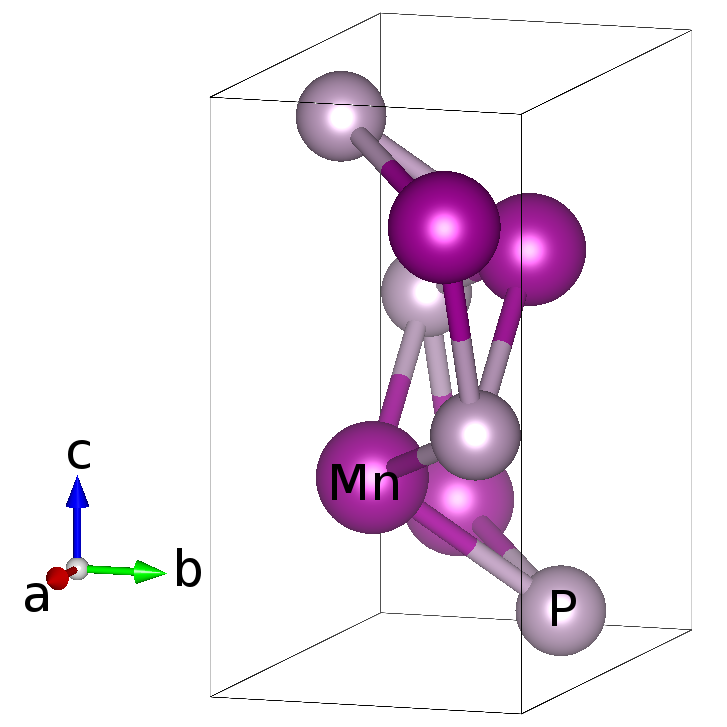}
    \caption{(Color online) Unit cell of MnP in the orthorhombic $Pnma$ structure.}\label{fig:uc}
\end{figure}

\begin{figure*}
    \includegraphics[]{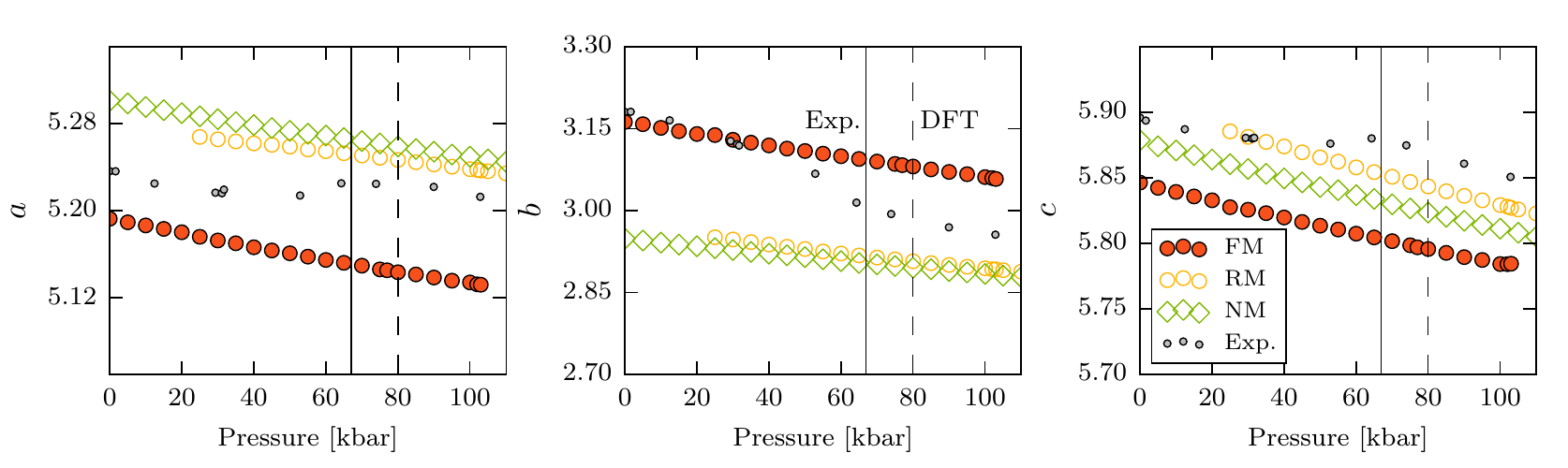}
    \caption{(Color online) Optimized lattice parameters as a function of applied pressure for the FM, the RM and the NM states (see text). The small gray dots are taken from the x-ray scattering experiment of Ref.~\cite{1511.09152}.}\label{fig:extendedlatpar}
\end{figure*}

To identify the magnetic structure of the non-collinear states we used the spin spiral method together with the full potential linearised augmented-plane wave (FP-LAPW) approach implemented in the Elk code \cite{elk}.
The energy differences involved in the stability of the helix states are extremely small and tight convergence is mandatory.
The basis set was defined with the following settings: rgkmax=8.5, gmaxvr=22, lmaxapw=10, nempty=12. The reader is referred to the Elk documentation for a detailed discussion of the parameters describing the basis set.
We selected a MP grid of $8\times12\times7$ points (or better) shifted by 0.5/$n_{i}$ ($i=x,y,z$) in each direction, which results in a total of 338 and 672 inequivalent points for the simulations with and without inversion symmetry respectively. Convergence was obtained with a smearing energy of 1~meV and using the Methfessel-Paxton scheme \cite{PhysRevB.40.3616}.
These settings guarantee total energy differences converged to better than 0.1 meV/atom.
All simulations were assumed to be converged when the total energy variation was less than 50 $\mu$eV and the root-mean-square change in Kohn-Sham potential was less than 3 $\mu$eV.
For the spin spirals simulations, when atomic positions were not available experimentally, the structures were predicted with DFT. To this aim we considered the FM ground state obtained from the experimental lattice parameters and we also included spin-orbit coupling.
We remind the reader that the spin orbit coupling is not present in spin spirals simulations.

Finally, to verify the accuracy of the planewave and pseudopotential approach, we compared the results obtained with QuantumESPRESSO and Elk. The energy difference between the ferromagnetic and the spinless solutions are -543~meV and -549~meV respectively for the planewave and the LAPW approaches. The band structures for the non magnetic ground state, compared in Fig.~\ref{fig:bands}, perfectly match.

\subsection{MnP unit cell and lattice constants}
The unit cell of MnP in the $Pnma$ setting is shown in Fig.~\ref{fig:uc}.
Both the Mn and the P atoms occupy occupy the $4c$ $(x,1/4,z)$ crystallographic positions with
$x_{\mathrm{Mn}}= 0.0049(2)$, $z_{\mathrm{Mn}}= 0.1965(2)$ and $x_{\mathrm{P}}= 0.1878(5)$,
$z_{\mathrm{P}}= 0.5686(5)$ \cite{Hulliger1968}.
These positions have been adopted for spin spiral calculations at ambient pressure. 

Figure~\ref{fig:extendedlatpar} shows instead evolution of the lattice parameters for the FM, the RM and the NM ground states.
The optimized lattice constant for the RM and the NM phases are similar except for the $c$ axis that is more expanded in the RM state than in the NM state.


\begin{thebibliography}{26}%
\makeatletter
\providecommand \@ifxundefined [1]{%
 \@ifx{#1\undefined}
}%
\providecommand \@ifnum [1]{%
 \ifnum #1\expandafter \@firstoftwo
 \else \expandafter \@secondoftwo
 \fi
}%
\providecommand \@ifx [1]{%
 \ifx #1\expandafter \@firstoftwo
 \else \expandafter \@secondoftwo
 \fi
}%
\providecommand \natexlab [1]{#1}%
\providecommand \enquote  [1]{``#1''}%
\providecommand \bibnamefont  [1]{#1}%
\providecommand \bibfnamefont [1]{#1}%
\providecommand \citenamefont [1]{#1}%
\providecommand \href@noop [0]{\@secondoftwo}%
\providecommand \href [0]{\begingroup \@sanitize@url \@href}%
\providecommand \@href[1]{\@@startlink{#1}\@@href}%
\providecommand \@@href[1]{\endgroup#1\@@endlink}%
\providecommand \@sanitize@url [0]{\catcode `\\12\catcode `\$12\catcode
  `\&12\catcode `\#12\catcode `\^12\catcode `\_12\catcode `\%12\relax}%
\providecommand \@@startlink[1]{}%
\providecommand \@@endlink[0]{}%
\providecommand \url  [0]{\begingroup\@sanitize@url \@url }%
\providecommand \@url [1]{\endgroup\@href {#1}{\urlprefix }}%
\providecommand \urlprefix  [0]{URL }%
\providecommand \Eprint [0]{\href }%
\providecommand \doibase [0]{http://dx.doi.org/}%
\providecommand \selectlanguage [0]{\@gobble}%
\providecommand \bibinfo  [0]{\@secondoftwo}%
\providecommand \bibfield  [0]{\@secondoftwo}%
\providecommand \translation [1]{[#1]}%
\providecommand \BibitemOpen [0]{}%
\providecommand \bibitemStop [0]{}%
\providecommand \bibitemNoStop [0]{.\EOS\space}%
\providecommand \EOS [0]{\spacefactor3000\relax}%
\providecommand \BibitemShut  [1]{\csname bibitem#1\endcsname}%
\let\auto@bib@innerbib\@empty
\bibitem [{\citenamefont {Moon}\ \emph {et~al.}(1981)\citenamefont {Moon},
  \citenamefont {Cable},\ and\ \citenamefont
  {Shapira}}]{Moon_Cable_Shapira_1981}%
  \BibitemOpen
  \bibfield  {author} {\bibinfo {author} {\bibfnamefont {R.}~\bibnamefont
  {Moon}}, \bibinfo {author} {\bibfnamefont {J.}~\bibnamefont {Cable}}, \ and\
  \bibinfo {author} {\bibfnamefont {Y.}~\bibnamefont {Shapira}},\ }\href@noop
  {} {\bibfield  {journal} {\bibinfo  {journal} {J. Appl. Phys.}\ }\textbf {\bibinfo {volume} {52}},\ \bibinfo
  {pages} {2025} (\bibinfo {year}
  {1981})}\BibitemShut {NoStop}%
\bibitem [{\citenamefont {Becerra}\ \emph {et~al.}(1980)\citenamefont
  {Becerra}, \citenamefont {Shapira}, \citenamefont {Oliveira},\ and\
  \citenamefont {Chang}}]{PhysRevLett.44.1692}%
  \BibitemOpen
  \bibfield  {author} {\bibinfo {author} {\bibfnamefont {C.~C.}\ \bibnamefont
  {Becerra}}, \bibinfo {author} {\bibfnamefont {Y.}~\bibnamefont {Shapira}},
  \bibinfo {author} {\bibfnamefont {N.~F.}\ \bibnamefont {Oliveira}}, \ and\
  \bibinfo {author} {\bibfnamefont {T.~S.}\ \bibnamefont {Chang}},\ }\href
  {\doibase 10.1103/PhysRevLett.44.1692} {\bibfield  {journal} {\bibinfo
  {journal} {Phys. Rev. Lett.}\ }\textbf {\bibinfo {volume} {44}},\ \bibinfo
  {pages} {1692} (\bibinfo {year} {1980})}\BibitemShut {NoStop}%
\bibitem [{\citenamefont {Cheng}\ \emph {et~al.}(2015)\citenamefont {Cheng},
  \citenamefont {Matsubayashi}, \citenamefont {Wu}, \citenamefont {Sun},
  \citenamefont {Lin}, \citenamefont {Luo},\ and\ \citenamefont
  {Uwatoko}}]{PhysRevLett.114.117001}%
  \BibitemOpen
  \bibfield  {author} {\bibinfo {author} {\bibfnamefont {J.-G.}\ \bibnamefont
  {Cheng}}, \bibinfo {author} {\bibfnamefont {K.}~\bibnamefont {Matsubayashi}},
  \bibinfo {author} {\bibfnamefont {W.}~\bibnamefont {Wu}}, \bibinfo {author}
  {\bibfnamefont {J.~P.}\ \bibnamefont {Sun}}, \bibinfo {author} {\bibfnamefont
  {F.~K.}\ \bibnamefont {Lin}}, \bibinfo {author} {\bibfnamefont {J.~L.}\
  \bibnamefont {Luo}}, \ and\ \bibinfo {author} {\bibfnamefont
  {Y.}~\bibnamefont {Uwatoko}},\ }\href {\doibase
  10.1103/PhysRevLett.114.117001} {\bibfield  {journal} {\bibinfo  {journal}
  {Phys. Rev. Lett.}\ }\textbf {\bibinfo {volume} {114}},\ \bibinfo {pages}
  {117001} (\bibinfo {year} {2015})}\BibitemShut {NoStop}%
\bibitem [{\citenamefont {Booth}\ and\ \citenamefont
  {Majetich}(2009)}]{1.3072022}%
  \BibitemOpen
  \bibfield  {author} {\bibinfo {author} {\bibfnamefont {R.~A.}\ \bibnamefont
  {Booth}}\ and\ \bibinfo {author} {\bibfnamefont {S.~A.}\ \bibnamefont
  {Majetich}},\ }\href {\doibase http://dx.doi.org/10.1063/1.3072022}
  {\bibfield  {journal} {\bibinfo  {journal} {Journal of Applied Physics}\
  }\textbf {\bibinfo {volume} {105}},\ \bibinfo {eid} {07A926} (\bibinfo {year}
  {2009})}\BibitemShut {NoStop}%
\bibitem [{\citenamefont {Chong}\ \emph {et~al.}(2016)\citenamefont {Chong},
  \citenamefont {Jiang}, \citenamefont {Zhou},\ and\ \citenamefont
  {Feng}}]{Chong2016}%
  \BibitemOpen
  \bibfield  {author} {\bibinfo {author} {\bibfnamefont {X.~Y.}\ \bibnamefont
  {Chong}}, \bibinfo {author} {\bibfnamefont {Y.}~\bibnamefont {Jiang}},
  \bibinfo {author} {\bibfnamefont {R.}~\bibnamefont {Zhou}}, \ and\ \bibinfo
  {author} {\bibfnamefont {J.}~\bibnamefont {Feng}},\ }\href
  {http://dx.doi.org/10.1038/srep21821} {\bibfield  {journal} {\bibinfo
  {journal} {Scientific Reports}\ }\textbf {\bibinfo {volume} {6}},\ \bibinfo
  {pages} {21821 EP} (\bibinfo {year} {2016})}\BibitemShut {NoStop}%
\bibitem [{\citenamefont {Huber}\ and\ \citenamefont
  {Ridgley}(1964)}]{PhysRev.135.A1033}%
  \BibitemOpen
  \bibfield  {author} {\bibinfo {author} {\bibfnamefont {E.~E.}\ \bibnamefont
  {Huber}}\ and\ \bibinfo {author} {\bibfnamefont {D.~H.}\ \bibnamefont
  {Ridgley}},\ }\href {\doibase 10.1103/PhysRev.135.A1033} {\bibfield
  {journal} {\bibinfo  {journal} {Phys. Rev.}\ }\textbf {\bibinfo {volume}
  {135}},\ \bibinfo {pages} {A1033} (\bibinfo {year} {1964})}\BibitemShut
  {NoStop}%
\bibitem [{\citenamefont {Felcher}(1966)}]{1.1708333}%
  \BibitemOpen
  \bibfield  {author} {\bibinfo {author} {\bibfnamefont {G.~P.}\ \bibnamefont
  {Felcher}},\ }\href {\doibase http://dx.doi.org/10.1063/1.1708333} {\bibfield
   {journal} {\bibinfo  {journal} {Journal of Applied Physics}\ }\textbf
  {\bibinfo {volume} {37}},\ \bibinfo {pages} {1056} (\bibinfo {year}
  {1966})}\BibitemShut {NoStop}%
\bibitem [{\citenamefont {Wang}\ \emph {et~al.}(2015)\citenamefont {Wang},
  \citenamefont {Feng}, \citenamefont {Cheng}, \citenamefont {Wu},
  \citenamefont {Luo},\ and\ \citenamefont {Rosenbaum}}]{1511.09152}%
  \BibitemOpen
  \bibfield  {author} {\bibinfo {author} {\bibfnamefont {Y.}~\bibnamefont
  {Wang}}, \bibinfo {author} {\bibfnamefont {Y.}~\bibnamefont {Feng}}, \bibinfo
  {author} {\bibfnamefont {J.~G.}\ \bibnamefont {Cheng}}, \bibinfo {author}
  {\bibfnamefont {W.}~\bibnamefont {Wu}}, \bibinfo {author} {\bibfnamefont
  {J.~L.}\ \bibnamefont {Luo}}, \ and\ \bibinfo {author} {\bibfnamefont
  {T.~F.}\ \bibnamefont {Rosenbaum}},\ }\href@noop {} {\enquote {\bibinfo
  {title} {Magnetic pitch as a tuning fork for superconductivity},}\ }
  (\bibinfo {year} {2015}),\ \Eprint {http://arxiv.org/abs/arXiv:1511.09152}
  {arXiv:1511.09152 [cond-mat]} \BibitemShut {NoStop}%
\bibitem [{\citenamefont {Hulliger}(1968)}]{Hulliger1968}%
  \BibitemOpen
  \bibfield  {author} {\bibinfo {author} {\bibfnamefont {F.}~\bibnamefont
  {Hulliger}},\ }\enquote {\bibinfo {title} {Structure and bonding},}\ \
  (\bibinfo  {publisher} {Springer Berlin Heidelberg},\ \bibinfo {address}
  {Berlin, Heidelberg},\ \bibinfo {year} {1968})\ Chap.\ \bibinfo {chapter}
  {Crystal chemistry of the chalcogenides and pnictides of the transition
  elements}, pp.\ \bibinfo {pages} {83--229}\BibitemShut {NoStop}%
\bibitem [{\citenamefont {Yamazaki}\ \emph {et~al.}(2014)\citenamefont
  {Yamazaki}, \citenamefont {Tabata}, \citenamefont {Waki}, \citenamefont
  {Sato}, \citenamefont {Matsuura}, \citenamefont {Ohoyama}, \citenamefont
  {Yokoyama},\ and\ \citenamefont {Nakamura}}]{JPSJ.83.054711}%
  \BibitemOpen
  \bibfield  {author} {\bibinfo {author} {\bibfnamefont {T.}~\bibnamefont
  {Yamazaki}}, \bibinfo {author} {\bibfnamefont {Y.}~\bibnamefont {Tabata}},
  \bibinfo {author} {\bibfnamefont {T.}~\bibnamefont {Waki}}, \bibinfo {author}
  {\bibfnamefont {T.~J.}\ \bibnamefont {Sato}}, \bibinfo {author}
  {\bibfnamefont {M.}~\bibnamefont {Matsuura}}, \bibinfo {author}
  {\bibfnamefont {K.}~\bibnamefont {Ohoyama}}, \bibinfo {author} {\bibfnamefont
  {M.}~\bibnamefont {Yokoyama}}, \ and\ \bibinfo {author} {\bibfnamefont
  {H.}~\bibnamefont {Nakamura}},\ }\href {\doibase 10.7566/JPSJ.83.054711}
  {\bibfield  {journal} {\bibinfo  {journal} {Journal of the Physical Society
  of Japan}\ }\textbf {\bibinfo {volume} {83}},\ \bibinfo {pages} {054711}
  (\bibinfo {year} {2014})} \BibitemShut {NoStop}%
\bibitem [{\citenamefont {Khasanov}\ \emph {et~al.}(2016)\citenamefont
  {Khasanov}, \citenamefont {Amato}, \citenamefont {Bonfà}, \citenamefont
  {Guguchia}, \citenamefont {Luetkens}, \citenamefont {Morenzoni},
  \citenamefont {Renzi},\ and\ \citenamefont {Zhigadlo}}]{1603.03367}%
  \BibitemOpen
  \bibfield  {author} {\bibinfo {author} {\bibfnamefont {R.}~\bibnamefont
  {Khasanov}}, \bibinfo {author} {\bibfnamefont {A.}~\bibnamefont {Amato}},
  \bibinfo {author} {\bibfnamefont {P.}~\bibnamefont {Bonfà}}, \bibinfo
  {author} {\bibfnamefont {Z.}~\bibnamefont {Guguchia}}, \bibinfo {author}
  {\bibfnamefont {H.}~\bibnamefont {Luetkens}}, \bibinfo {author}
  {\bibfnamefont {E.}~\bibnamefont {Morenzoni}}, \bibinfo {author}
  {\bibfnamefont {R.~D.}\ \bibnamefont {Renzi}}, \ and\ \bibinfo {author}
  {\bibfnamefont {N.~D.}\ \bibnamefont {Zhigadlo}},\ }\href@noop {} {\enquote
  {\bibinfo {title} {High pressure magnetic state of mnp probed by means of
  muon-spin rotation},}\ } (\bibinfo {year} {2016}),\ \Eprint
  {http://arxiv.org/abs/arXiv:1603.03367} {arXiv:1603.03367 [cond-mat]}
  \BibitemShut {NoStop}%
\bibitem [{\citenamefont {Matsuda}\ \emph {et~al.}(2016)\citenamefont
  {Matsuda}, \citenamefont {Ye}, \citenamefont {Dissanayake}, \citenamefont
  {Cheng}, \citenamefont {Chi}, \citenamefont {Ma}, \citenamefont {Zhou},
  \citenamefont {Yan}, \citenamefont {Kasamatsu}, \citenamefont {Sugino},
  \citenamefont {Kato}, \citenamefont {Matsubayashi}, \citenamefont {Okada},\
  and\ \citenamefont {Uwatoko}}]{PhysRevB.93.100405}%
  \BibitemOpen
  \bibfield  {author} {\bibinfo {author} {\bibfnamefont {M.}~\bibnamefont
  {Matsuda}}, \bibinfo {author} {\bibfnamefont {F.}~\bibnamefont {Ye}},
  \bibinfo {author} {\bibfnamefont {S.~E.}\ \bibnamefont {Dissanayake}},
  \bibinfo {author} {\bibfnamefont {J.-G.}\ \bibnamefont {Cheng}}, \bibinfo
  {author} {\bibfnamefont {S.}~\bibnamefont {Chi}}, \bibinfo {author}
  {\bibfnamefont {J.}~\bibnamefont {Ma}}, \bibinfo {author} {\bibfnamefont
  {H.~D.}\ \bibnamefont {Zhou}}, \bibinfo {author} {\bibfnamefont {J.-Q.}\
  \bibnamefont {Yan}}, \bibinfo {author} {\bibfnamefont {S.}~\bibnamefont
  {Kasamatsu}}, \bibinfo {author} {\bibfnamefont {O.}~\bibnamefont {Sugino}},
  \bibinfo {author} {\bibfnamefont {T.}~\bibnamefont {Kato}}, \bibinfo {author}
  {\bibfnamefont {K.}~\bibnamefont {Matsubayashi}}, \bibinfo {author}
  {\bibfnamefont {T.}~\bibnamefont {Okada}}, \ and\ \bibinfo {author}
  {\bibfnamefont {Y.}~\bibnamefont {Uwatoko}},\ }\href {\doibase
  10.1103/PhysRevB.93.100405} {\bibfield  {journal} {\bibinfo  {journal} {Phys.
  Rev. B}\ }\textbf {\bibinfo {volume} {93}},\ \bibinfo {pages} {100405}
  (\bibinfo {year} {2016})}\BibitemShut {NoStop}%
\bibitem [{\citenamefont {Gercsi}\ and\ \citenamefont
  {Sandeman}(2010)}]{PhysRevB.81.224426}%
  \BibitemOpen
  \bibfield  {author} {\bibinfo {author} {\bibfnamefont {Z.}~\bibnamefont
  {Gercsi}}\ and\ \bibinfo {author} {\bibfnamefont {K.~G.}\ \bibnamefont
  {Sandeman}},\ }\href {\doibase 10.1103/PhysRevB.81.224426} {\bibfield
  {journal} {\bibinfo  {journal} {Phys. Rev. B}\ }\textbf {\bibinfo {volume}
  {81}},\ \bibinfo {pages} {224426} (\bibinfo {year} {2010})}\BibitemShut
  {NoStop}%
\bibitem [{\citenamefont {Continenza}\ \emph {et~al.}(2001)\citenamefont
  {Continenza}, \citenamefont {Picozzi}, \citenamefont {Geng},\ and\
  \citenamefont {Freeman}}]{PhysRevB.64.085204}%
  \BibitemOpen
  \bibfield  {author} {\bibinfo {author} {\bibfnamefont {A.}~\bibnamefont
  {Continenza}}, \bibinfo {author} {\bibfnamefont {S.}~\bibnamefont {Picozzi}},
  \bibinfo {author} {\bibfnamefont {W.~T.}\ \bibnamefont {Geng}}, \ and\
  \bibinfo {author} {\bibfnamefont {A.~J.}\ \bibnamefont {Freeman}},\ }\href
  {\doibase 10.1103/PhysRevB.64.085204} {\bibfield  {journal} {\bibinfo
  {journal} {Phys. Rev. B}\ }\textbf {\bibinfo {volume} {64}},\ \bibinfo
  {pages} {085204} (\bibinfo {year} {2001})}\BibitemShut {NoStop}%
\bibitem [{\citenamefont {Okabayashi}\ \emph {et~al.}(2004)\citenamefont
  {Okabayashi}, \citenamefont {Tanaka}, \citenamefont {Hashimoto},
  \citenamefont {Fujimori}, \citenamefont {Ono}, \citenamefont {Okusawa},\ and\
  \citenamefont {Komatsubara}}]{PhysRevB.69.132411}%
  \BibitemOpen
  \bibfield  {author} {\bibinfo {author} {\bibfnamefont {J.}~\bibnamefont
  {Okabayashi}}, \bibinfo {author} {\bibfnamefont {K.}~\bibnamefont {Tanaka}},
  \bibinfo {author} {\bibfnamefont {M.}~\bibnamefont {Hashimoto}}, \bibinfo
  {author} {\bibfnamefont {A.}~\bibnamefont {Fujimori}}, \bibinfo {author}
  {\bibfnamefont {K.}~\bibnamefont {Ono}}, \bibinfo {author} {\bibfnamefont
  {M.}~\bibnamefont {Okusawa}}, \ and\ \bibinfo {author} {\bibfnamefont
  {T.}~\bibnamefont {Komatsubara}},\ }\href {\doibase
  10.1103/PhysRevB.69.132411} {\bibfield  {journal} {\bibinfo  {journal} {Phys.
  Rev. B}\ }\textbf {\bibinfo {volume} {69}},\ \bibinfo {pages} {132411}
  (\bibinfo {year} {2004})}\BibitemShut {NoStop}%
\bibitem [{\citenamefont {Giannozzi}\ \emph {et~al.}(2009)\citenamefont
  {Giannozzi}, \citenamefont {Baroni}, \citenamefont {Bonini}, \citenamefont
  {Calandra}, \citenamefont {Car}, \citenamefont {Cavazzoni}, \citenamefont
  {Ceresoli}, \citenamefont {Chiarotti}, \citenamefont {Cococcioni},
  \citenamefont {Dabo}, \citenamefont {{Dal Corso}}, \citenamefont {ano~de
  Gironcoli}, \citenamefont {Fabris}, \citenamefont {Fratesi}, \citenamefont
  {Gebauer}, \citenamefont {Gerstmann}, \citenamefont {Gougoussis},
  \citenamefont {Kokalj}, \citenamefont {Lazzeri}, \citenamefont
  {Martin-Samos}, \citenamefont {Marzari}, \citenamefont {Mauri}, \citenamefont
  {an~d Stefano~Paolini}, \citenamefont {Pasquarello}, \citenamefont
  {Paulatto}, \citenamefont {Sbraccia}, \citenamefont {Scandolo}, \citenamefont
  {Sclauzero}, \citenamefont {Seitsonen}, \citenamefont {Smogunov},
  \citenamefont {Umari},\ and\ \citenamefont {Wentzcovitch}}]{QE-2009}%
  \BibitemOpen
  \bibfield  {author} {\bibinfo {author} {\bibfnamefont {P.}~\bibnamefont
  {Giannozzi}}, \bibinfo {author} {\bibfnamefont {S.}~\bibnamefont {Baroni}},
  \bibinfo {author} {\bibfnamefont {N.}~\bibnamefont {Bonini}}, \bibinfo
  {author} {\bibfnamefont {M.}~\bibnamefont {Calandra}}, \bibinfo {author}
  {\bibfnamefont {R.}~\bibnamefont {Car}}, \bibinfo {author} {\bibfnamefont
  {C.}~\bibnamefont {Cavazzoni}}, \bibinfo {author} {\bibfnamefont
  {D.}~\bibnamefont {Ceresoli}}, \bibinfo {author} {\bibfnamefont {G.~L.}\
  \bibnamefont {Chiarotti}}, \bibinfo {author} {\bibfnamefont {M.}~\bibnamefont
  {Cococcioni}}, \bibinfo {author} {\bibfnamefont {I.}~\bibnamefont {Dabo}},
  \bibinfo {author} {\bibfnamefont {A.}~\bibnamefont {{Dal Corso}}}, \bibinfo
  {author} {\bibfnamefont {S.}~\bibnamefont {ano~de Gironcoli}}, \bibinfo
  {author} {\bibfnamefont {S.}~\bibnamefont {Fabris}}, \bibinfo {author}
  {\bibfnamefont {G.}~\bibnamefont {Fratesi}}, \bibinfo {author} {\bibfnamefont
  {R.}~\bibnamefont {Gebauer}}, \bibinfo {author} {\bibfnamefont
  {U.}~\bibnamefont {Gerstmann}}, \bibinfo {author} {\bibfnamefont
  {C.}~\bibnamefont {Gougoussis}}, \bibinfo {author} {\bibfnamefont
  {A.}~\bibnamefont {Kokalj}}, \bibinfo {author} {\bibfnamefont
  {M.}~\bibnamefont {Lazzeri}}, \bibinfo {author} {\bibfnamefont
  {L.}~\bibnamefont {Martin-Samos}}, \bibinfo {author} {\bibfnamefont
  {N.}~\bibnamefont {Marzari}}, \bibinfo {author} {\bibfnamefont
  {F.}~\bibnamefont {Mauri}}, \bibinfo {author} {\bibfnamefont {R.~M.}\
  \bibnamefont {an~d Stefano~Paolini}}, \bibinfo {author} {\bibfnamefont
  {A.}~\bibnamefont {Pasquarello}}, \bibinfo {author} {\bibfnamefont
  {L.}~\bibnamefont {Paulatto}}, \bibinfo {author} {\bibfnamefont
  {C.}~\bibnamefont {Sbraccia}}, \bibinfo {author} {\bibfnamefont
  {S.}~\bibnamefont {Scandolo}}, \bibinfo {author} {\bibfnamefont
  {G.}~\bibnamefont {Sclauzero}}, \bibinfo {author} {\bibfnamefont {A.~P.}\
  \bibnamefont {Seitsonen}}, \bibinfo {author} {\bibfnamefont {A.}~\bibnamefont
  {Smogunov}}, \bibinfo {author} {\bibfnamefont {P.}~\bibnamefont {Umari}}, \
  and\ \bibinfo {author} {\bibfnamefont {R.~M.}\ \bibnamefont {Wentzcovitch}},\
  }\href {http://www.quantum-espresso.org} {\bibfield  {journal} {\bibinfo
  {journal} {Journal of Physics: Condensed Matter}\ }\textbf {\bibinfo {volume}
  {21}},\ \bibinfo {pages} {395502 (19pp)} (\bibinfo {year}
  {2009})}\BibitemShut {NoStop}%
\bibitem [{\citenamefont {Dewhurst}\ \emph {et~al.}(2016)\citenamefont
  {Dewhurst}, \citenamefont {Sharma}, \citenamefont {Nordström}, \citenamefont
  {Cricchio}, \citenamefont {Grånäs}, \citenamefont {Gross}, \citenamefont
  {Ambrosch-Draxl}, \citenamefont {Persson}, \citenamefont {Bultmark},
  \citenamefont {Brouder}, \citenamefont {Armiento}, \citenamefont
  {Chizmeshya}, \citenamefont {Anderson}, \citenamefont {Nekrasov},
  \citenamefont {Wagner}, \citenamefont {Kalarasse}, \citenamefont {Spitaler},
  \citenamefont {Pittalis}, \citenamefont {Lathiotakis}, \citenamefont
  {Burnus}, \citenamefont {Sagmeister}, \citenamefont {Meisenbichler},
  \citenamefont {Lebègue}, \citenamefont {Zhang}, \citenamefont {Körmann},
  \citenamefont {Baranov}, \citenamefont {Kozhevnikov}, \citenamefont
  {Suehara}, \citenamefont {Essenberger}, \citenamefont {Sanna}, \citenamefont
  {McQueen}, \citenamefont {Baldsiefen}, \citenamefont {Blaber}, \citenamefont
  {Filanovich}, \citenamefont {Björkman}, \citenamefont {Stankovski},
  \citenamefont {Goraus}, \citenamefont {Meinert}, \citenamefont {Rohr},
  \citenamefont {Nazarov}, \citenamefont {Krieger}, \citenamefont {Floyd},
  \citenamefont {Davydov}, \citenamefont {Eich}, \citenamefont {Castro},
  \citenamefont {Kitahara}, \citenamefont {Glasbrenner}, \citenamefont
  {Bussmann}, \citenamefont {Mazin}, \citenamefont {Verstraete}, \citenamefont
  {Ernsting}, \citenamefont {Dugdale}, \citenamefont {Elliott}, \citenamefont
  {Dulak}, \citenamefont {Livas}, \citenamefont {Cottenier},\ and\
  \citenamefont {Shinohara}}]{elk}%
  \BibitemOpen
  \bibfield  {author} {\bibinfo {author} {\bibfnamefont {K.}~\bibnamefont
  {Dewhurst}}, \bibinfo {author} {\bibfnamefont {S.}~\bibnamefont {Sharma}},
  \bibinfo {author} {\bibfnamefont {L.}~\bibnamefont {Nordström}}, \bibinfo
  {author} {\bibfnamefont {F.}~\bibnamefont {Cricchio}}, \bibinfo {author}
  {\bibfnamefont {O.}~\bibnamefont {Grånäs}}, \bibinfo {author}
  {\bibfnamefont {H.}~\bibnamefont {Gross}}, \bibinfo {author} {\bibfnamefont
  {C.}~\bibnamefont {Ambrosch-Draxl}}, \bibinfo {author} {\bibfnamefont
  {C.}~\bibnamefont {Persson}}, \bibinfo {author} {\bibfnamefont
  {F.}~\bibnamefont {Bultmark}}, \bibinfo {author} {\bibfnamefont
  {C.}~\bibnamefont {Brouder}}, \bibinfo {author} {\bibfnamefont
  {R.}~\bibnamefont {Armiento}}, \bibinfo {author} {\bibfnamefont
  {A.}~\bibnamefont {Chizmeshya}}, \bibinfo {author} {\bibfnamefont
  {P.}~\bibnamefont {Anderson}}, \bibinfo {author} {\bibfnamefont
  {I.}~\bibnamefont {Nekrasov}}, \bibinfo {author} {\bibfnamefont
  {F.}~\bibnamefont {Wagner}}, \bibinfo {author} {\bibfnamefont
  {F.}~\bibnamefont {Kalarasse}}, \bibinfo {author} {\bibfnamefont
  {J.}~\bibnamefont {Spitaler}}, \bibinfo {author} {\bibfnamefont
  {S.}~\bibnamefont {Pittalis}}, \bibinfo {author} {\bibfnamefont
  {N.}~\bibnamefont {Lathiotakis}}, \bibinfo {author} {\bibfnamefont
  {T.}~\bibnamefont {Burnus}}, \bibinfo {author} {\bibfnamefont
  {S.}~\bibnamefont {Sagmeister}}, \bibinfo {author} {\bibfnamefont
  {C.}~\bibnamefont {Meisenbichler}}, \bibinfo {author} {\bibfnamefont
  {S.}~\bibnamefont {Lebègue}}, \bibinfo {author} {\bibfnamefont
  {Y.}~\bibnamefont {Zhang}}, \bibinfo {author} {\bibfnamefont
  {F.}~\bibnamefont {Körmann}}, \bibinfo {author} {\bibfnamefont
  {A.}~\bibnamefont {Baranov}}, \bibinfo {author} {\bibfnamefont
  {A.}~\bibnamefont {Kozhevnikov}}, \bibinfo {author} {\bibfnamefont
  {S.}~\bibnamefont {Suehara}}, \bibinfo {author} {\bibfnamefont
  {F.}~\bibnamefont {Essenberger}}, \bibinfo {author} {\bibfnamefont
  {A.}~\bibnamefont {Sanna}}, \bibinfo {author} {\bibfnamefont
  {T.}~\bibnamefont {McQueen}}, \bibinfo {author} {\bibfnamefont
  {T.}~\bibnamefont {Baldsiefen}}, \bibinfo {author} {\bibfnamefont
  {M.}~\bibnamefont {Blaber}}, \bibinfo {author} {\bibfnamefont
  {A.}~\bibnamefont {Filanovich}}, \bibinfo {author} {\bibfnamefont
  {T.}~\bibnamefont {Björkman}}, \bibinfo {author} {\bibfnamefont
  {M.}~\bibnamefont {Stankovski}}, \bibinfo {author} {\bibfnamefont
  {J.}~\bibnamefont {Goraus}}, \bibinfo {author} {\bibfnamefont
  {M.}~\bibnamefont {Meinert}}, \bibinfo {author} {\bibfnamefont
  {D.}~\bibnamefont {Rohr}}, \bibinfo {author} {\bibfnamefont {V.}~\bibnamefont
  {Nazarov}}, \bibinfo {author} {\bibfnamefont {K.}~\bibnamefont {Krieger}},
  \bibinfo {author} {\bibfnamefont {P.}~\bibnamefont {Floyd}}, \bibinfo
  {author} {\bibfnamefont {A.}~\bibnamefont {Davydov}}, \bibinfo {author}
  {\bibfnamefont {F.}~\bibnamefont {Eich}}, \bibinfo {author} {\bibfnamefont
  {A.~R.}\ \bibnamefont {Castro}}, \bibinfo {author} {\bibfnamefont
  {K.}~\bibnamefont {Kitahara}}, \bibinfo {author} {\bibfnamefont
  {J.}~\bibnamefont {Glasbrenner}}, \bibinfo {author} {\bibfnamefont
  {K.}~\bibnamefont {Bussmann}}, \bibinfo {author} {\bibfnamefont
  {I.}~\bibnamefont {Mazin}}, \bibinfo {author} {\bibfnamefont
  {M.}~\bibnamefont {Verstraete}}, \bibinfo {author} {\bibfnamefont
  {D.}~\bibnamefont {Ernsting}}, \bibinfo {author} {\bibfnamefont
  {S.}~\bibnamefont {Dugdale}}, \bibinfo {author} {\bibfnamefont
  {P.}~\bibnamefont {Elliott}}, \bibinfo {author} {\bibfnamefont
  {M.}~\bibnamefont {Dulak}}, \bibinfo {author} {\bibfnamefont {J.~A.~F.}\
  \bibnamefont {Livas}}, \bibinfo {author} {\bibfnamefont {S.}~\bibnamefont
  {Cottenier}}, \ and\ \bibinfo {author} {\bibfnamefont {Y.}~\bibnamefont
  {Shinohara}},\ }\href@noop {} {\enquote {\bibinfo {title} {{Elk code version
  3.3.17}},}\ }\bibinfo {howpublished} {\url{http://elk.sf.net}} (\bibinfo
  {year} {2016}),\ \bibinfo {note} {[Online; accessed
  16-March-2016]}\BibitemShut {NoStop}%
\bibitem [{\citenamefont {Perdew}\ \emph {et~al.}(1997)\citenamefont {Perdew},
  \citenamefont {Burke},\ and\ \citenamefont
  {Ernzerhof}}]{PhysRevLett.78.1396}%
  \BibitemOpen
  \bibfield  {author} {\bibinfo {author} {\bibfnamefont {J.~P.}\ \bibnamefont
  {Perdew}}, \bibinfo {author} {\bibfnamefont {K.}~\bibnamefont {Burke}}, \
  and\ \bibinfo {author} {\bibfnamefont {M.}~\bibnamefont {Ernzerhof}},\ }\href
  {\doibase 10.1103/PhysRevLett.78.1396} {\bibfield  {journal} {\bibinfo
  {journal} {Phys. Rev. Lett.}\ }\textbf {\bibinfo {volume} {78}},\ \bibinfo
  {pages} {1396} (\bibinfo {year} {1997})}\BibitemShut {NoStop}%
\bibitem [{\citenamefont {Methfessel}\ and\ \citenamefont
  {Paxton}(1989)}]{PhysRevB.40.3616}%
  \BibitemOpen
  \bibfield  {author} {\bibinfo {author} {\bibfnamefont {M.}~\bibnamefont
  {Methfessel}}\ and\ \bibinfo {author} {\bibfnamefont {A.~T.}\ \bibnamefont
  {Paxton}},\ }\href {\doibase 10.1103/PhysRevB.40.3616} {\bibfield  {journal}
  {\bibinfo  {journal} {Phys. Rev. B}\ }\textbf {\bibinfo {volume} {40}},\
  \bibinfo {pages} {3616} (\bibinfo {year} {1989})}\BibitemShut {NoStop}%
\bibitem [{\citenamefont {Monkhorst}\ and\ \citenamefont
  {Pack}(1976)}]{PhysRevB.13.5188}%
  \BibitemOpen
  \bibfield  {author} {\bibinfo {author} {\bibfnamefont {H.~J.}\ \bibnamefont
  {Monkhorst}}\ and\ \bibinfo {author} {\bibfnamefont {J.~D.}\ \bibnamefont
  {Pack}},\ }\href {\doibase 10.1103/PhysRevB.13.5188} {\bibfield  {journal}
  {\bibinfo  {journal} {Phys. Rev. B}\ }\textbf {\bibinfo {volume} {13}},\
  \bibinfo {pages} {5188} (\bibinfo {year} {1976})}\BibitemShut {NoStop}%
\bibitem [{\citenamefont {Marvel}(2016)}]{sssp}%
  \BibitemOpen
  \bibfield  {author} {\bibinfo {author} {\bibnamefont {Marvel}},\ }\href@noop
  {} {\enquote {\bibinfo {title} {{Standard Solid State Pseudopotentials
  (SSSP)}},}\ }\bibinfo {howpublished} {\url{http://materialscloud.org/sssp/}}
  (\bibinfo {year} {2016}),\ \bibinfo {note} {[Online; accessed
  16-March-2016]}\BibitemShut {NoStop}%
\bibitem [{\citenamefont {Lejaeghere}\ \emph {et~al.}(2016)\citenamefont
  {Lejaeghere}, \citenamefont {Bihlmayer}, \citenamefont {Bj{\"o}rkman},
  \citenamefont {Blaha}, \citenamefont {Bl{\"u}gel}, \citenamefont {Blum},
  \citenamefont {Caliste}, \citenamefont {Castelli}, \citenamefont {Clark},
  \citenamefont {Dal~Corso}, \citenamefont {de~Gironcoli}, \citenamefont
  {Deutsch}, \citenamefont {Dewhurst}, \citenamefont {Di~Marco}, \citenamefont
  {Draxl}, \citenamefont {Du{\l}ak}, \citenamefont {Eriksson}, \citenamefont
  {Flores-Livas}, \citenamefont {Garrity}, \citenamefont {Genovese},
  \citenamefont {Giannozzi}, \citenamefont {Giantomassi}, \citenamefont
  {Goedecker}, \citenamefont {Gonze}, \citenamefont {Gr{\r a}n{\"a}s},
  \citenamefont {Gross}, \citenamefont {Gulans}, \citenamefont {Gygi},
  \citenamefont {Hamann}, \citenamefont {Hasnip}, \citenamefont {Holzwarth},
  \citenamefont {Iu{\c s}an}, \citenamefont {Jochym}, \citenamefont {Jollet},
  \citenamefont {Jones}, \citenamefont {Kresse}, \citenamefont {Koepernik},
  \citenamefont {K{\"u}{\c c}{\"u}kbenli}, \citenamefont {Kvashnin},
  \citenamefont {Locht}, \citenamefont {Lubeck}, \citenamefont {Marsman},
  \citenamefont {Marzari}, \citenamefont {Nitzsche}, \citenamefont
  {Nordstr{\"o}m}, \citenamefont {Ozaki}, \citenamefont {Paulatto},
  \citenamefont {Pickard}, \citenamefont {Poelmans}, \citenamefont {Probert},
  \citenamefont {Refson}, \citenamefont {Richter}, \citenamefont {Rignanese},
  \citenamefont {Saha}, \citenamefont {Scheffler}, \citenamefont {Schlipf},
  \citenamefont {Schwarz}, \citenamefont {Sharma}, \citenamefont {Tavazza},
  \citenamefont {Thunstr{\"o}m}, \citenamefont {Tkatchenko}, \citenamefont
  {Torrent}, \citenamefont {Vanderbilt}, \citenamefont {van Setten},
  \citenamefont {Van~Speybroeck}, \citenamefont {Wills}, \citenamefont {Yates},
  \citenamefont {Zhang},\ and\ \citenamefont {Cottenier}}]{Lejaeghereaad3000}%
  \BibitemOpen
  \bibfield  {author} {\bibinfo {author} {\bibfnamefont {K.}~\bibnamefont
  {Lejaeghere}}, \bibinfo {author} {\bibfnamefont {G.}~\bibnamefont
  {Bihlmayer}}, \bibinfo {author} {\bibfnamefont {T.}~\bibnamefont
  {Bj{\"o}rkman}}, \bibinfo {author} {\bibfnamefont {P.}~\bibnamefont {Blaha}},
  \bibinfo {author} {\bibfnamefont {S.}~\bibnamefont {Bl{\"u}gel}}, \bibinfo
  {author} {\bibfnamefont {V.}~\bibnamefont {Blum}}, \bibinfo {author}
  {\bibfnamefont {D.}~\bibnamefont {Caliste}}, \bibinfo {author} {\bibfnamefont
  {I.~E.}\ \bibnamefont {Castelli}}, \bibinfo {author} {\bibfnamefont {S.~J.}\
  \bibnamefont {Clark}}, \bibinfo {author} {\bibfnamefont {A.}~\bibnamefont
  {Dal~Corso}}, \bibinfo {author} {\bibfnamefont {S.}~\bibnamefont
  {de~Gironcoli}}, \bibinfo {author} {\bibfnamefont {T.}~\bibnamefont
  {Deutsch}}, \bibinfo {author} {\bibfnamefont {J.~K.}\ \bibnamefont
  {Dewhurst}}, \bibinfo {author} {\bibfnamefont {I.}~\bibnamefont {Di~Marco}},
  \bibinfo {author} {\bibfnamefont {C.}~\bibnamefont {Draxl}}, \bibinfo
  {author} {\bibfnamefont {M.}~\bibnamefont {Du{\l}ak}}, \bibinfo {author}
  {\bibfnamefont {O.}~\bibnamefont {Eriksson}}, \bibinfo {author}
  {\bibfnamefont {J.~A.}\ \bibnamefont {Flores-Livas}}, \bibinfo {author}
  {\bibfnamefont {K.~F.}\ \bibnamefont {Garrity}}, \bibinfo {author}
  {\bibfnamefont {L.}~\bibnamefont {Genovese}}, \bibinfo {author}
  {\bibfnamefont {P.}~\bibnamefont {Giannozzi}}, \bibinfo {author}
  {\bibfnamefont {M.}~\bibnamefont {Giantomassi}}, \bibinfo {author}
  {\bibfnamefont {S.}~\bibnamefont {Goedecker}}, \bibinfo {author}
  {\bibfnamefont {X.}~\bibnamefont {Gonze}}, \bibinfo {author} {\bibfnamefont
  {O.}~\bibnamefont {Gr{\r a}n{\"a}s}}, \bibinfo {author} {\bibfnamefont
  {E.~K.~U.}\ \bibnamefont {Gross}}, \bibinfo {author} {\bibfnamefont
  {A.}~\bibnamefont {Gulans}}, \bibinfo {author} {\bibfnamefont
  {F.}~\bibnamefont {Gygi}}, \bibinfo {author} {\bibfnamefont {D.~R.}\
  \bibnamefont {Hamann}}, \bibinfo {author} {\bibfnamefont {P.~J.}\
  \bibnamefont {Hasnip}}, \bibinfo {author} {\bibfnamefont {N.~A.~W.}\
  \bibnamefont {Holzwarth}}, \bibinfo {author} {\bibfnamefont {D.}~\bibnamefont
  {Iu{\c s}an}}, \bibinfo {author} {\bibfnamefont {D.~B.}\ \bibnamefont
  {Jochym}}, \bibinfo {author} {\bibfnamefont {F.}~\bibnamefont {Jollet}},
  \bibinfo {author} {\bibfnamefont {D.}~\bibnamefont {Jones}}, \bibinfo
  {author} {\bibfnamefont {G.}~\bibnamefont {Kresse}}, \bibinfo {author}
  {\bibfnamefont {K.}~\bibnamefont {Koepernik}}, \bibinfo {author}
  {\bibfnamefont {E.}~\bibnamefont {K{\"u}{\c c}{\"u}kbenli}}, \bibinfo
  {author} {\bibfnamefont {Y.~O.}\ \bibnamefont {Kvashnin}}, \bibinfo {author}
  {\bibfnamefont {I.~L.~M.}\ \bibnamefont {Locht}}, \bibinfo {author}
  {\bibfnamefont {S.}~\bibnamefont {Lubeck}}, \bibinfo {author} {\bibfnamefont
  {M.}~\bibnamefont {Marsman}}, \bibinfo {author} {\bibfnamefont
  {N.}~\bibnamefont {Marzari}}, \bibinfo {author} {\bibfnamefont
  {U.}~\bibnamefont {Nitzsche}}, \bibinfo {author} {\bibfnamefont
  {L.}~\bibnamefont {Nordstr{\"o}m}}, \bibinfo {author} {\bibfnamefont
  {T.}~\bibnamefont {Ozaki}}, \bibinfo {author} {\bibfnamefont
  {L.}~\bibnamefont {Paulatto}}, \bibinfo {author} {\bibfnamefont {C.~J.}\
  \bibnamefont {Pickard}}, \bibinfo {author} {\bibfnamefont {W.}~\bibnamefont
  {Poelmans}}, \bibinfo {author} {\bibfnamefont {M.~I.~J.}\ \bibnamefont
  {Probert}}, \bibinfo {author} {\bibfnamefont {K.}~\bibnamefont {Refson}},
  \bibinfo {author} {\bibfnamefont {M.}~\bibnamefont {Richter}}, \bibinfo
  {author} {\bibfnamefont {G.-M.}\ \bibnamefont {Rignanese}}, \bibinfo {author}
  {\bibfnamefont {S.}~\bibnamefont {Saha}}, \bibinfo {author} {\bibfnamefont
  {M.}~\bibnamefont {Scheffler}}, \bibinfo {author} {\bibfnamefont
  {M.}~\bibnamefont {Schlipf}}, \bibinfo {author} {\bibfnamefont
  {K.}~\bibnamefont {Schwarz}}, \bibinfo {author} {\bibfnamefont
  {S.}~\bibnamefont {Sharma}}, \bibinfo {author} {\bibfnamefont
  {F.}~\bibnamefont {Tavazza}}, \bibinfo {author} {\bibfnamefont
  {P.}~\bibnamefont {Thunstr{\"o}m}}, \bibinfo {author} {\bibfnamefont
  {A.}~\bibnamefont {Tkatchenko}}, \bibinfo {author} {\bibfnamefont
  {M.}~\bibnamefont {Torrent}}, \bibinfo {author} {\bibfnamefont
  {D.}~\bibnamefont {Vanderbilt}}, \bibinfo {author} {\bibfnamefont {M.~J.}\
  \bibnamefont {van Setten}}, \bibinfo {author} {\bibfnamefont
  {V.}~\bibnamefont {Van~Speybroeck}}, \bibinfo {author} {\bibfnamefont
  {J.~M.}\ \bibnamefont {Wills}}, \bibinfo {author} {\bibfnamefont {J.~R.}\
  \bibnamefont {Yates}}, \bibinfo {author} {\bibfnamefont {G.-X.}\ \bibnamefont
  {Zhang}}, \ and\ \bibinfo {author} {\bibfnamefont {S.}~\bibnamefont
  {Cottenier}},\ }\href {\doibase 10.1126/science.aad3000} {\bibfield
  {journal} {\bibinfo  {journal} {Science}\ }\textbf {\bibinfo {volume} {351}}
  (\bibinfo {year} {2016}),\ 10.1126/science.aad3000},\ \Eprint
  {http://arxiv.org/abs/http://science.sciencemag.org/content/351/6280/aad3000.full.pdf}
  {http://science.sciencemag.org/content/351/6280/aad3000.full.pdf}
  \BibitemShut {NoStop}%
\bibitem [{\citenamefont {Bl\"ochl}(1994)}]{PhysRevB.50.17953}%
  \BibitemOpen
  \bibfield  {author} {\bibinfo {author} {\bibfnamefont {P.~E.}\ \bibnamefont
  {Bl\"ochl}},\ }\href {\doibase 10.1103/PhysRevB.50.17953} {\bibfield
  {journal} {\bibinfo  {journal} {Phys. Rev. B}\ }\textbf {\bibinfo {volume}
  {50}},\ \bibinfo {pages} {17953} (\bibinfo {year} {1994})}\BibitemShut
  {NoStop}%
\bibitem [{Note1()}]{Note1}%
  \BibitemOpen
  \bibinfo {note} {The FM state is the closest collinear equivalent of the
  helical state which is observed below 50K at ambient pressure.}\BibitemShut
  {Stop}%
\bibitem [{Note2()}]{Note2}%
  \BibitemOpen
  \bibinfo {note} {We cannot exclude the possibility of a phase transition to a
  lattice structure with lower symmetry at $p<280$~kbar}\BibitemShut {NoStop}%
\bibitem [{\citenamefont {ichiro Yano}\ \emph {et~al.}(2013)\citenamefont
  {ichiro Yano}, \citenamefont {Itoh}, \citenamefont {Yokoo}, \citenamefont
  {Satoh}, \citenamefont {Kawana}, \citenamefont {Kousaka}, \citenamefont
  {Akimitsu},\ and\ \citenamefont {Endoh}}]{Yano201333}%
  \BibitemOpen
  \bibfield  {author} {\bibinfo {author} {\bibfnamefont {S.}~\bibnamefont
  {ichiro Yano}}, \bibinfo {author} {\bibfnamefont {S.}~\bibnamefont {Itoh}},
  \bibinfo {author} {\bibfnamefont {T.}~\bibnamefont {Yokoo}}, \bibinfo
  {author} {\bibfnamefont {S.}~\bibnamefont {Satoh}}, \bibinfo {author}
  {\bibfnamefont {D.}~\bibnamefont {Kawana}}, \bibinfo {author} {\bibfnamefont
  {Y.}~\bibnamefont {Kousaka}}, \bibinfo {author} {\bibfnamefont
  {J.}~\bibnamefont {Akimitsu}}, \ and\ \bibinfo {author} {\bibfnamefont
  {Y.}~\bibnamefont {Endoh}},\ }\href {\doibase
  http://dx.doi.org/10.1016/j.jmmm.2013.07.019} {\bibfield  {journal} {\bibinfo
   {journal} {Journal of Magnetism and Magnetic Materials}\ }\textbf {\bibinfo
  {volume} {347}},\ \bibinfo {pages} {33 } (\bibinfo {year}
  {2013})}\BibitemShut {NoStop}%
\bibitem [{Note3()}]{Note3}%
  \BibitemOpen
  \bibinfo {note} {To obtain accurate results, we decided to fix the lattice
  parameters to the known experimental values\cite {1511.09152}. Since, to the
  best of our knowledge, the atomic positions as a function of the pressure are
  still unknown, we optimized them in the ferromagnetic state including also
  the spin orbit coupling.}\BibitemShut {Stop}%
\end{thebibliography}
\end{document}